\documentclass[12pt,twoside]{article}
\usepackage{jeffe,graphicx,url,fullpage,enumerate,verbatim}
\newtheorem{lemma}{Lemma}[section]
\newtheorem{theorem}[lemma]{Theorem}
\newtheorem{corollary}[lemma]{Corollary}
\begin{document}

\pagestyle{myheadings}
\markboth{Jeff Erickson and Sariel Har-Peled}
         {Optimally Cutting a Surface into a Disk}
\urldef{\paperurl}\url{http://www.cs.uiuc.edu/~jeffe/pubs/schema.html}

\begin{titlepage}

\title{Optimally Cutting a Surface into a Disk\thanks{%
	A preliminary version of this paper was presented at the 18th
        Annual ACM Symposium on Computational Geometry~\cite{eh-ocsd-02}.
        See \paperurl\ for the most recent version of this paper.}}

\author{Jeff Erickson\thanks{%
	    Partially supported by a Sloan Fellowship, NSF CAREER
	    award CCR-0093348, and NSF ITR grant DMR-0121695.}
	\qquad
	Sariel Har-Peled\thanks{%
	    Partially supported by NSF CAREER award CCR-0132901.}
	\\[2ex]
	\normalsize
	\begin{tabular}{c}
	    University of Illinois at Urbana-Champaign\\
	    \url{{jeffe,sariel}@cs.uiuc.edu}\\
	    \url{http://www.cs.uiuc.edu/~{jeffe,sariel}}
	\end{tabular}}

\date{Submitted to \textsl{Discrete \& Computational Geometry}: \today}

\maketitle

\begin{abstract}
We consider the problem of cutting a set of edges on a polyhedral
manifold surface, possibly with boundary, to obtain a single
topological disk, minimizing either the total number of cut edges or
their total length.  We show that this problem is NP-hard, even for
manifolds without boundary and for punctured spheres.  We also
describe an algorithm with running time $n^{O(g+k)}$, where $n$ is the
combinatorial complexity, $g$ is the genus, and $k$ is the number of
boundary components of the input surface.  Finally, we describe a
greedy algorithm that outputs a $O(\log^2 g)$-approximation of the
minimum cut graph in $O(g^2n\log n)$ time.
\end{abstract}

\thispagestyle{empty}
\setcounter{page}{0}
\end{titlepage}

\def\M{\mathcal{M}}
\def\MM{\overline{\M}}

% ----------------------------------------------------------------------
\section{Introduction}  

Several applications of three-dimensional surfaces require information
about the underlying topological structure in addition to the
geometry.  In some cases, we wish to simplify the surface topology, to
facilitate algorithms that can be performed only if the surface is a
topological disk.

Applications when this is important include surface parameterization
\cite{f-psast-97,ss-spmtf-00} and texture mapping~\cite{bvig-psfnd-91,
pb-stmss-00}.  In the texture mapping problem, we wish to find a
continuous and invertible mapping from the texture, usually a
two-dimensional rectangular image, to the surface.  Unfortunately, if
the surface is not a topological disk, no such map exists.  In such a
case, the only feasible solution is to cut the surface so that it
becomes a topological disk.  (Haker \etal\ \cite{hatksh-csptm-00}
present an algorithm for directly texture mapping models with the
topology of a sphere, where the texture is also embedded on a sphere.)
Of course, when cutting the surface, one would like to find the best
possible cut under various considerations.  For example, one might
want to cut the surface so that the resulting surface can be textured
mapped with minimum distortion \cite{f-psast-97, ggh-gi-02,
ss-spmtf-00}.  To our knowledge, all previous approaches for this
cutting problem either rely on heuristics with no quality guarantees
\cite{ggh-gi-02, s-stsrp-02, sf-c3dfo-02} or require the user to
perform this cutting beforehand~\cite{f-psast-97, pb-stmss-00}.

Lazarus \etal~\cite{lpvv-ccpso-01} presented and implemented two
algorithms for computing a canonical polygonal schema of an orientable
surface of complexity $n$ and with genus $g$, in time $O(gn)$,
simplifying an earlier algorithm of Vegter and Yap~\cite{vy-cccs-90}.
Computing such a schema requires finding $2g$ cycles, all passing
through a common basepoint in $\M$, such that cutting along those
cycles breaks~$\M$ into a topological disk.  Since these cycles must
share a common point, it is easy to find examples where the overall
size of those cycles is $\Omega(gn)$.  Furthermore, those cycles share
several edges and are visually unsatisfying.

\begin{figure}
\centerline{\includegraphics[height=2.25in]{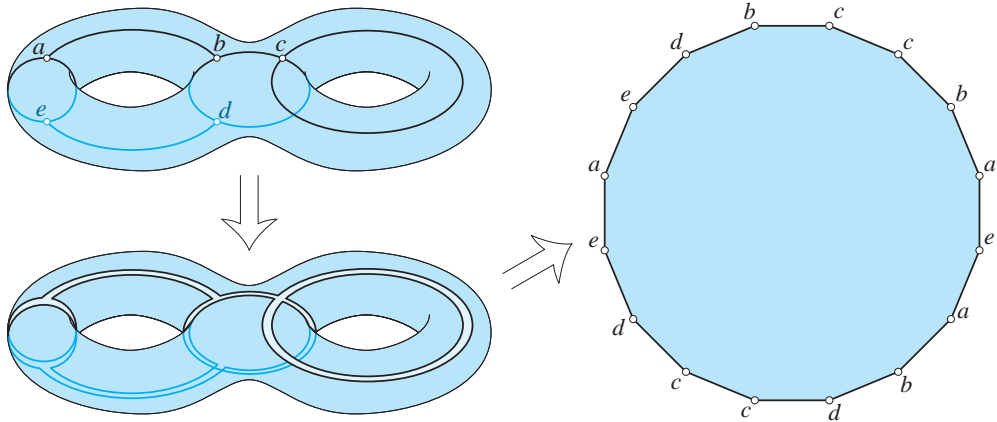}}
\caption{A cut graph for a two-holed torus and its induced
   (non-canonical) polygonal schema.}
\label{F:cutgraph}
\end{figure}

For most applications, computing a canonical schema is overkill.  It
is usually sufficient to find a collection of edges whose removal
transforms the surface into a topological disk.  We call such a set of
edges a \emph{cut graph}; see Figure \ref{F:cutgraph} for an example.
Cut graphs have several advantages.  First, they are compact.
Trivially, any cut graph contains at most $n$ edges of the surface
mesh, much less than any canonical schema in the worst case, although
we expect it to be much smaller in practice.  Second, it is quite easy
to construct a cut graph for an arbitrary polyhedral surface in $O(n)$
time, using a breadth-first search of the dual
graph~\cite{ds-ntcps-95}, or simply taking a maximal set of edges
whose complement is connected~\cite{lpvv-ccpso-01}.  Finally, the cut
graph has an extremely simple structure: a tree with $O(g)$ additional
edges.  As such, it should be easier to manipulate algorithmically
than other representations.  For example, Dey and Schipper
\cite{ds-ntcps-95} describe fast algorithms to determine whether a
curve is contractible, or two curves are homotopic, using an arbitrary
cut graph instead of a canonical schema.

In this paper, we investigate the question of how find the ``best''
such cutting of a surface, restricting ourselves to cuts along the
edges of the given mesh.  Specifically, we want to find the smallest
subset of edges of a polyhedral manifold surface $\M$, possibly with
boundary, such that cutting along those edges transforms $\M$ into a
topological disk.  We also consider the weighted version of this
problem, where each edge has an arbitrary non-negative weight and we
want to minimize the total weight of the cut graph.  The most natural
weight of an edge is its Euclidean length, but we could also assign
weights to take problem-specific considerations into account.  For
example, if we want to compute a texture mapping for a specific
viewpoint, we could make visible edges more expensive, so that the
minimum cut graph would minimize the number of visible edges used in
the cuts.  Our algorithms do not require the edge weights to satisfy
the triangle inequality.

We show that the minimum cut graph of any polyhedral manifold~$\M$
with genus~$g$ and $k$~boundary components can be computed in
$n^{O(g+k)}$ time.  We also show that the problem is NP-hard in
general, even if $g$ or $k$ is fixed.  Finally, we present a simple
and efficient greedy approximation algorithm for this problem.  Our
algorithm outputs a cut graph whose weight is a factor $O(\log^2 g)$
larger than optimal, in $O(g^2 n \log n)$ time.%
\footnote{To simplify notation, we define $\log x = \max\set{1,
\ceil{\log_2 x}}$.}
If $g=0$, the approximation factor is exactly $2$.  As a tool in our
approximation algorithm, we also describe efficient algorithms to
compute shortest and nearly-shortest nontrivial cycles in a manifold;
we believe these algorithms are of independent interest.

% ----------------------------------------------------------------------
\section{Background}
\label{S:back}

Before presenting our new results, we review several useful notions
from topology and describe related results in more detail.  We refer
the interested reader to Hatcher~\cite{h-at-01},
Munkres~\cite{m-t-00}, or Stillwell~\cite{s-ctcgt-93} for further
topological background and more formal definitions.  For related
computational results, see the recent surveys by Dey, Edelsbrunner,
and Guha~\cite{deg-ct-99} and Vegter~\cite{v-ct-97}.

\subsection{Topology}

A \emph{$2$-manifold with boundary} is a set $\M$ such that every
point $x\in \M$ lies in a neighborhood homeomorphic to either the
plane $\Real^2$ or a closed halfplane.  The points with only halfplane
neighborhoods constitute the \emph{boundary} of $\M$; the boundary
consists of zero or more disjoint circles.  This paper will consider
only \emph{compact} manifolds, where every infinite sequence of points
has a convergent subsequence.

The \emph{genus} of a 2-manifold $\M$ is the maximum number of
disjoint non-separating cycles $\gamma_1, \gamma_2, \dots, \gamma_g$
in $\M$; that is, $\gamma_i\cap\gamma_j = \varnothing$ for all $i$ and
$j$, and $\M\setminus(\gamma_1\cup\cdots\cup\gamma_g)$ is connected.
For example, a sphere and a disc have genus $0$, a torus and a
\Mobius\ strip have genus~$1$, and a Klein bottle has genus~$2$.

A manifold is \emph{orientable} if it has two distinct sides, and
\emph{non-orientable} if it has only one side.  Although many
geometric applications use only orientable $2$-manifolds (primarily
because non-orientable manifolds without boundary cannot be embedded
in $\Real^3$ without self-intersections) our results will apply to
non-orientable manifolds as well.  Every (compact, connected)
$2$-manifold with boundary is characterized by its orientability, its
genus $g$, and the number $k$ of boundary components~\cite{fw-czp-99}.

A \emph{polyhedral} $2$-manifold is constructed by gluing closed
simple polygons edge-to-edge into a \emph{cell complex}: the
intersection of any two polygons is either empty, a vertex of both, or
an edge of both.  We refer to the component polygons has
\emph{facets}.  (Since the facets are closed, every polyhedral
manifold is compact.)  For any polyhedral manifold $\M$, the number of
vertices and facets, minus the number of edges, is the \emph{Euler
characteristic} $\chi$ of~$\M$.  Euler's formula~\cite{e-spefv-01}
implies that $\chi$ is an invariant of the underlying manifold,
independent of any particular polyhedral representation; $\chi =
2-2g-k$ if the manifold is orientable, and $\chi = 2-g-k$ if the
manifold is non-orientable.  Euler's formula implies that if $\M$ has
$v$ vertices, then $\M$ has at most $3v-6+6g$ edges and at most
$2v-4+4g-k$ facets, with equality for orientable manifolds where every
facet and boundary circle is a triangle.  We let $n \le 6v - 10 + 10g
- k$ denote the total number of facets, edges, and vertices in $\M$.

The \emph{$1$-skeleton} $\M_1$ of a polyhedral manifold $\M$ is the
graph consisting of its vertices and edges.  We define a \emph{cut
graph} $G$ of $\M$ as a subgraph of $\M_1$ such that $\M\setminus G$
is homeomorphic to a disk.%
\footnote{Cut graphs are generalizations of the \emph{cut locus} of a
manifold $\M$, which is essentially the geodesic medial axis of
a single point.}
The disk $\M\setminus G$ is known as a \emph{polygonal schema} of
$\M$.  Each edge of $G$ appears twice on the boundary of polygonal
schema $\M\setminus G$, and we can obtain $\M$ by gluing together
these corresponding boundary edges.  Finding a cut graph of $\M$ with
minimum total length is clearly equivalent to to finding a polygonal
schema of $\M$ with minimum perimeter.

Any $2$-manifold has a so-called \emph{canonical} polygonal schema,
whose combinatorial structure depends only on the genus $g$, the
number of boundary components~$k$, and whether the manifold is
orientable.\footnote{Actually, there are several different ways to
define canonical schemata; the one described here is merely the most
common.  For example, the canonical schema for an oriented surface
without boundary could also be labeled $x_1, x_2, \dots, x_{2g},
\bar{x}_1, \bar{x}_2, \dots, \bar{x}_{2g}$.}  The canonical schema of
an orientable manifold is a ${(4g+3k)}$-gon with successive edges
labeled
\[
        x_1, y_1, \bar{x}_1, \bar{y}_1,\,
        \dots,
        x_g, y_g, \bar{x}_g, \bar{y}_g,\,
        z_1, e_1, \bar{z}_1, \,
        \dots,
        z_k, e_k, \bar{z}_k;
\]
for a non-orientable manifold, the canonical schema is a $(2g+3k)$-gon
with edge labels
\[
        x_1, x_1,\,
        \dots,
        x_g, x_g,\,
        z_1, e_1, \bar{z}_1, \,
        \dots,
        z_k, e_k, \bar{z}_k.
\]
Every pair of corresponding edges $x$ and $\bar{x}$ is oriented in
opposite directions.  Gluing together corresponding pairs in the
indicated directions recovers the original manifold, with the
unmatched edges $e_i$ forming the boundary circles.  For a manifold
$\M$ without boundary, a \emph{reduced} polygonal schema is one where
all the vertices are glued into a single point in~$\M$; canonical
schemata of manifolds without boundary are reduced.  We emphasize that
the polygonal schemata constructed by our algorithms are neither
necessarily canonical nor necessarily reduced.

\subsection{Previous and Related Results}

Dey and Schipper~\cite{ds-ntcps-95} describe an algorithm to construct
a reduced, but not necessarily canonical, polygonal schema for any
triangulated orientable manifold without boundary in $O(n)$ time.
Essentially, their algorithm constructs an arbitrary cut graph~$G$ by
depth-first search, and and then shrinks a spanning tree of $G$ to a
single point.  (See also Dey and Guha~\cite{dg-tcs-99}.)

Vegter and Yap~\cite{vy-cccs-90} developed an algorithm to construct a
canonical schema in optimal $O(gn)$ time and space.  Two simpler
algorithms with the same running time were later developed by Lazarus
\etal~\cite{lpvv-ccpso-01}.  The ``edges'' of the polygonal schemata
produced by all these algorithms are (possibly overlapping) paths in
the $1$-skeleton of the input manifold.  We will modify one of the
algorithms of Lazarus \etal\ to construct short nontrivial cycles and
cut graphs.

Very recently, Colin~de~Verdi\'ere and Lazarus consider the problem of
optimizing canonical polygonal schemata~\cite{cl-osflo-02}.  Given a
canonical polygonal schema for a triangulated oriented manifold $\M$,
their algorithm constructs the shortest canonical schema in the same
homotopy class.  Surprisingly (in light of our Theorem~\ref{Th:hard})
their algorithm runs in polynomial time under some mild assumptions
about the input.  As a byproduct, they also obtain a polynomial-time
algorithm to construct the minimum-length simple loop homotopic to a
given path.

Surface parameterization is an extremely active area of research,
thanks to numerous applications such as texture mapping, remeshing,
compression, and morphing.  For a sample of recent results, see
\cite{amd-igr-02, eddhl-maam-95, f-psast-97, sf-c3dfo-02, ggh-gi-02,
lsscd-mmaps-98, lprm-lscma-02, s-stsrp-02, scgl-bdpmp-02, ss-spmtf-00,
zkk-tmusf-01} and references therein.  In most of these works,
surfaces of high genus are parameterized by cutting them into several
(possibly overlapping) patches, each homeomorphic to a disk, each with
a separate parameterization.  A recent exception is the work of Gu
\etal~\cite{ggh-gi-02}, which computes an initial cut graph in
$O(n\log n)$ time by running a shortest path algorithm on the dual of
the manifold mesh, starting from an arbitrary seed triangle.
Essentially the same algorithm was independently proposed by Steiner
and Fischer \cite{sf-c3dfo-02}.  Once a surface has been cut into a
disk (or several disks), further (topologically trivial) cuts are
usually necessary to reduce distortion~\cite{ggh-gi-02, s-stsrp-02,
scgl-bdpmp-02}.  Many of these algorithms include heuristics to
minimize the lengths of the cuts in addition to the distortion of the
parameterization \cite{ggh-gi-02, lprm-lscma-02, s-stsrp-02}, but none
with theoretical guarantees.

All of our algorithms are ultimately based on Dijkstra's single-source
shortest path algorithm~\cite{d-ntpcg-59,t-dsna-83}.  Many previous
results have used Dijkstra's algorithm or one of its continuous
generalizations \cite{kab-fspsu-95, mmp-dgp-87, t-eagot-95} to
discover interesting topological structures in $2$-manifolds, such as
cut graphs~\cite{ggh-gi-02, sf-c3dfo-02}, small handles (`topological
noise')~\cite{gw-tnr-01}, texture atlases~\cite{lprm-lscma-02},
contour trees~\cite{ae-amatm-98, lv-lsdpo-99}, and Reeb
graphs~\cite{hskk-tmfas-01, sf-c3dfo-02}.

% ----------------------------------------------------------------------
\section{Computing Minimum Cut Graphs is NP-Hard}
\label{S:nphard}

In this section, we prove that finding a minimum cut graph of a
triangulated manifold is NP-hard.  We consider two versions of the
problem.  In the \emph{weighted} case, the manifold is assumed to be a
polyhedral surface in $\Real^3$ and we want to compute the cut graph
whose total Euclidean length is as small as possible.  In the
\emph{unweighted} case, we are compute the cut graph with the minimum
number of edges; the geometry of the manifold is ignored entirely.

Both reductions are from the \emph{rectilinear Steiner tree} problem:
Given a set $P$ of $n$ points from a $m\times m$ square grid in the
plane, find the shortest connected set of horizontal and vertical line
segments that contains every point in $P$.  This problem is NP-hard,
even if $m$~is bounded by a polynomial in~$n$~\cite{gj-rstpi-77}.  Our
reduction uses the \emph{Hanan grid} of the points, which is obtained
by drawing horizontal and vertical lines through each point, clipped
to the bounding box of the points.  At least one rectilinear Steiner
tree of the points is a subset of the Hanan grid~\cite{h-sprd-66}.

\begin{theorem}
\label{Th:hard}
Computing the length of the minimum (weighted or unweighted) cut graph
of a triangulated punctured sphere is NP-hard.
\end{theorem}

\begin{proof}
First consider the weighted case, where the weight assigned to each
edge of the manifold is its Euclidean length.  Let $P$ be a set of $n$
points in the plane, with integer coordinates between $1$ and~$m$.  We
construct a punctured sphere in $O(n^2)$ time as follows.  Assume that
$P$ lies on the $x y$-plane in $\Real^3$.  We modify the Hanan grid of
$P$ by replacing each terminal with a square of width $1/2n$, rotated
$45$ degrees so that its vertices lie on the neighboring edges.  These
squares will form the punctures.  We then attach a \emph{basin} under
each face $f$ of the modified Hanan grid, by joining the boundary of
$f$ to a slightly scaled copy of $f$ on the plane $z = -n^2$.  We also
attach a basin of depth $n^2+1$ to the boundary of the entire modified
Hanan grid.  The side facets of each basin are trapezoids.  The basins
are tapered so that adjacent basins intersect only on the modified
Hanan grid.  Triangulating this surface arbitrarily, we obtain a
polyhedral sphere $\M$ with $n$ punctures and overall complexity
$O(n^2)$.  See Figure~\ref{F:basins}.

\begin{figure}
\centering\footnotesize\sf
\begin{tabular}{c@{\qquad}c@{\qquad}c}
    \includegraphics[height=1.5in]{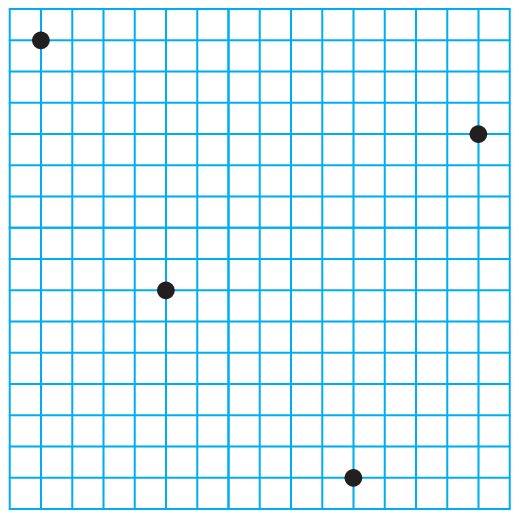} &
    \includegraphics[height=1.5in]{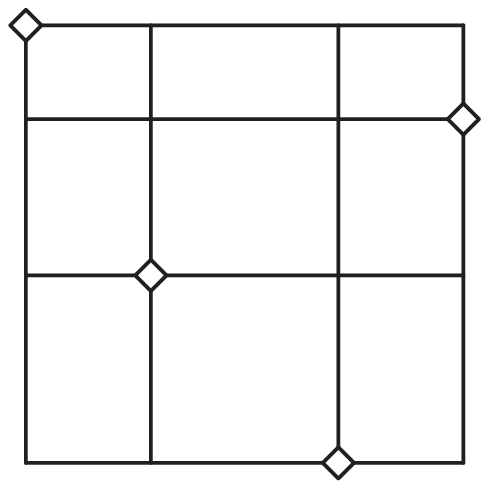} &
    \includegraphics[height=2.5in]{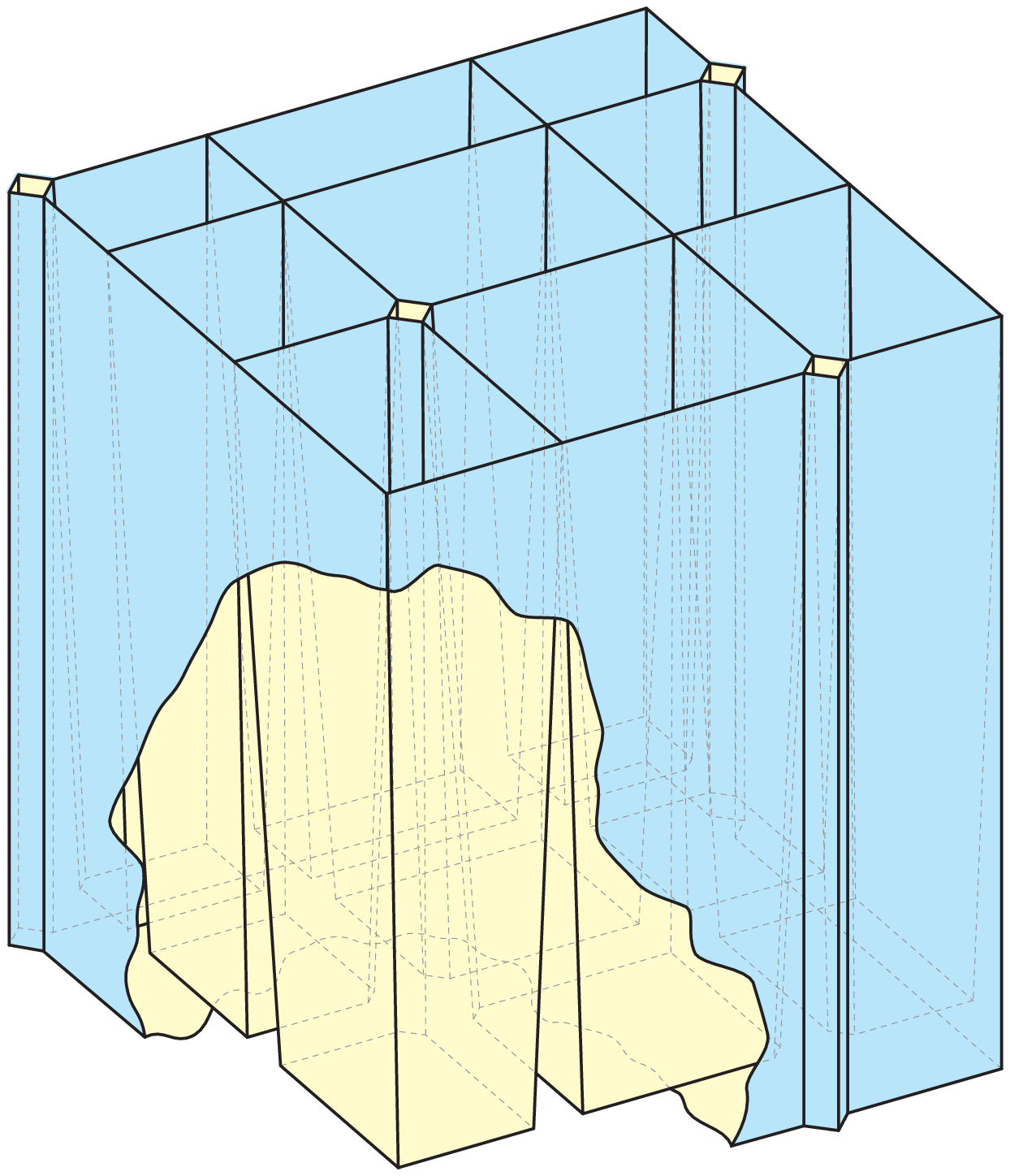} \\
    (a) & (b) & (c)
\end{tabular}
\caption{(a) A set of integer points.  (b) The modified Hanan grid.
(c) A cut-away view of the resulting punctured sphere.}
\label{F:basins}
\end{figure}

Let $G^*$ be a minimum weighted cut graph of $\M$.  We easily observe
that $G^*$ contains only ``long'' edges from the modified Hanan grid
and contains at least one vertex of every puncture.  Thus, the edges
of $G^*$ are in one-to-one correspondence with the edge of a
rectilinear Steiner tree of $P$.

For the unweighted case, we modify the original $m\times m$ integer
grid instead of the Hanan grid.  To create a punctured sphere, we
replace each terminal point with a small diamond as above.  We then
fill in each modified grid cell with a triangulation, chosen so that
the shortest path between any two points on the boundary of any cell
stays on the boundary of that cell; this requires a constant number of
triangles per cell.  The resulting manifold $\M'$ has complexity
$O(m^2)$.  By induction, the shortest path between any two points on
the modified grid lies entirely on the grid.  Thus, any minimal
unweighted cut graph of $\M'$ contains only edges from the modified
grid.  It follows that if the minimum unweighted cut graph of $\M'$
has $r$ edges, the length of any rectilinear Steiner tree of $P$ is
exactly~$r$.
\end{proof}

We can easily generalize the previous proof to manifolds with higher
genus, with or without boundary, oriented or not, by attaching small
triangulated tori or cross-caps to any subset of punctures.

\begin{theorem}
Computing the length of the minimum (weighted or unweighted) cut
graph of a triangulated manifold with boundary, with any fixed
genus or with any fixed number of boundary components, is NP-hard.
\end{theorem}

% ----------------------------------------------------------------------
\section{Computing Minimum Cut Graphs Anyway}
\label{S:algorithm}

We now describe an algorithm to compute the minimum cut graph of a
polyhedral manifold in $n^{O(g+k)}$ time.  For manifolds with constant
Euler characteristic, our algorithm runs in polynomial time.

Our algorithm is based on the following characterization of the
minimum cut graph as the union of shortest paths.  A \emph{branch
point} of a cut graph is any vertex with degree greater than~$2$.  A
simple path in a cut graph from one branch point or boundary point to
another, with no branch points in its interior, is called a \emph{cut
path}.

\begin{lemma}
\label{L:tight}
Let $\M$ be a polyhedral $2$-manifold, possibly with boundary, and
let $G^*$ be a minimum cut graph of~$\M$.  Any cut path in $G^*$
can be decomposed into two equal-length shortest paths in~$\M_1$.
\end{lemma}

\begin{figure}[t]
    \centerline{\includegraphics[width=\textwidth]{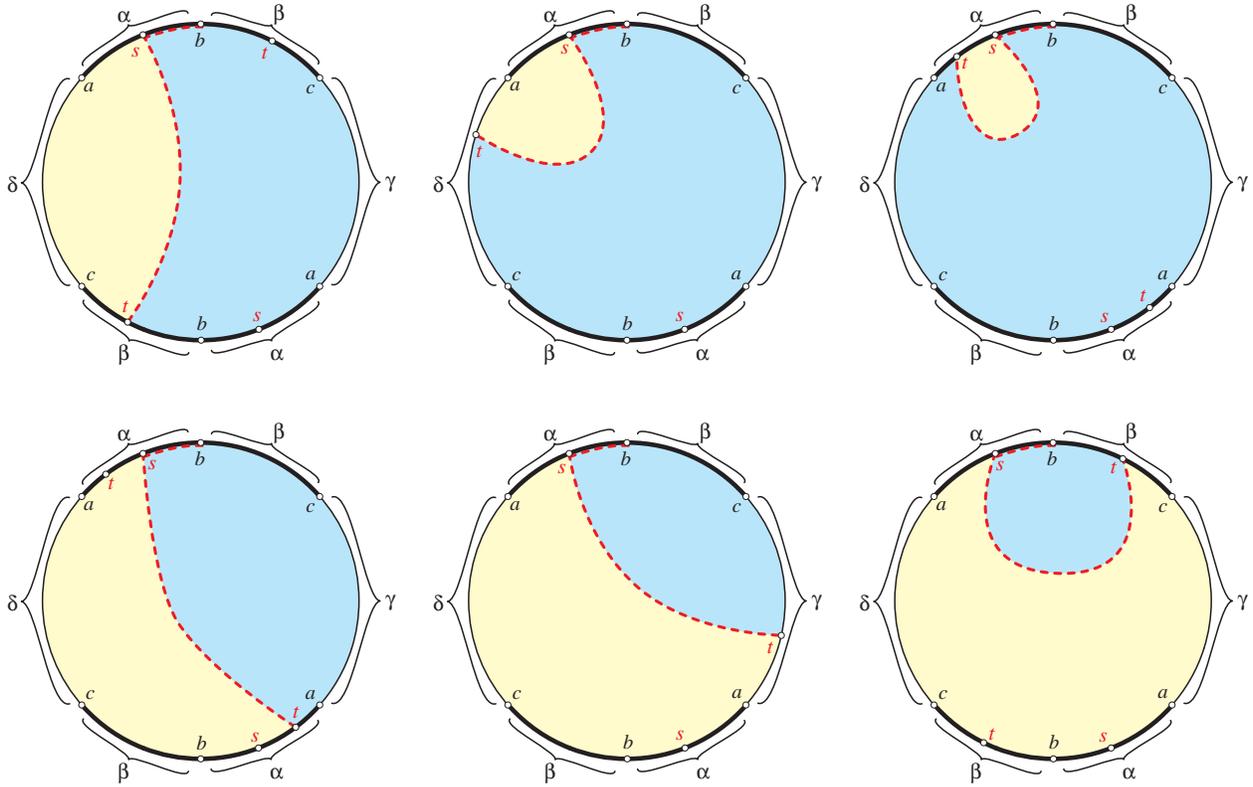}}
    \caption{Six additional cases for the proof of Lemma \ref{L:tight}
       for non-orientable manifolds; all other cases are reflections
       or rotations of these.  In each case, some subpath
       of~$\alpha$~or~$\beta$ appears on the boundary of both
       sub-disks.}
    \label{F:cases2}
\end{figure}

\begin{proof}
Let $G$ be an arbitrary cut graph of $\M$, and consider a cut path
between two (not necessarily distinct) branch points $a$ and $c$ of
$G$.  Let $b$ be the midpoint of this path, and let $\alpha$ and
$\beta$ denote the subpaths from $b$ to~$a$ and from $b$ to~$c$,
respectively.  Note that $b$ may lie in the interior of an edge of
$\M_1$.  Finally, suppose $\alpha$ is \emph{not} the shortest path
from $b$ to $a$ in~$\M_1$.  To prove the lemma, it suffices to show
that $G$ is not the shortest cut graph of $\M$.

\def\x#1{\widetilde{\smash{#1}\vphantom{t}}}
 
Let $\alpha'$ be the true shortest path from $b$ to $a$.  Clearly,
$\alpha'$ is not contained in $G$.  Walking along $\alpha'$ from $b$
to~$a$, let $s$ be the first vertex whose following edge is not
in~$G$, and let~$t$ be the first vertex in $G$ whose preceding edge is
not in~$G$.  (Note that $s$ and $t$ may be joined by a single edge in
$\M\setminus G$.)  Finally, let $\sigma'\subset\alpha'$ be the true
shortest path from $s$ to~$t$.  Equivalently, $\sigma'$ is the first
maximal subpath of $\alpha'$ whose interior lies in $\M\setminus G$.
See Figure~\ref{F:cutpath}.

\begin{figure}[th]
\centerline{\includegraphics[height=4in]{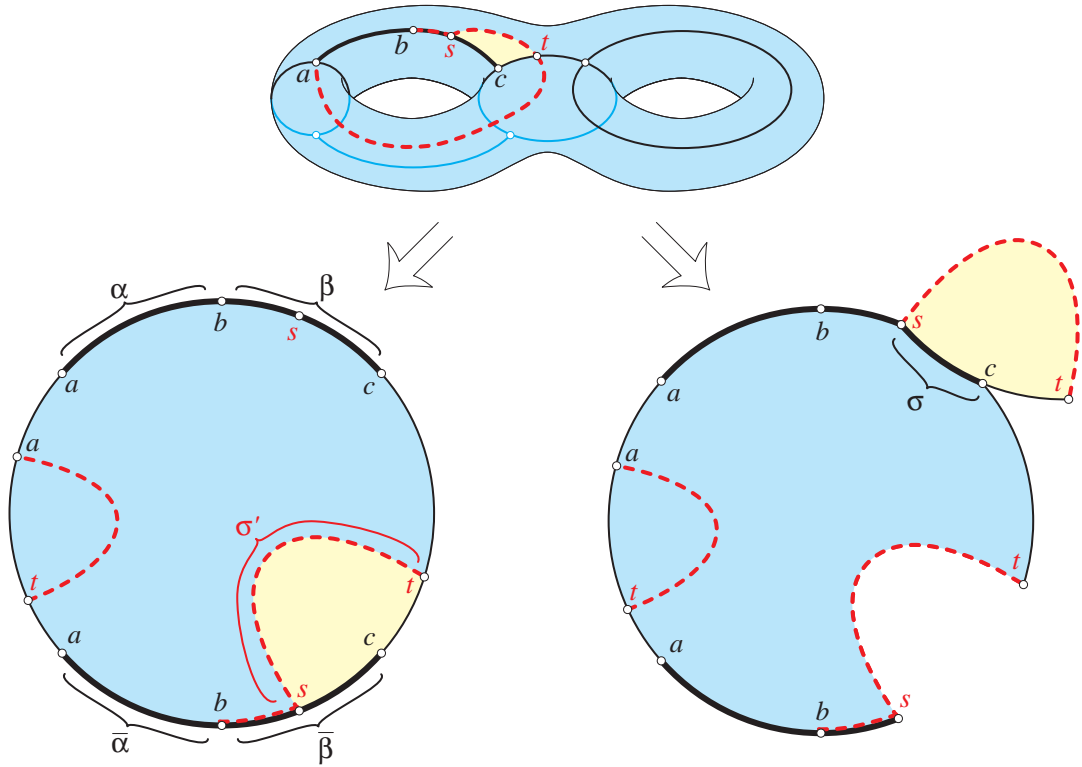}}
\caption{If the dashed path from $a$ to $b$ is shorter than $\alpha$,
then the cut graph can be shortened by cutting along $\sigma'$ and
regluing along $\sigma$.}
\label{F:cutpath}
\end{figure}

The subpath $\sigma'$ cuts $\M\setminus G$ into two smaller disks.  We
claim that some subpath $\sigma$ of either $\alpha$ or $\beta$ appears
on the boundary of both disks and is longer than $\sigma'$.  Our claim
implies that cutting $\M\setminus G$ along $\sigma'$ and regluing the
pieces along $\sigma$ gives us a new polygonal schema with smaller
perimeter, and thus a new cut graph shorter than $G$.  See
Figure~\ref{F:cutpath} for an example.

We prove our claim by exhaustive case analysis.  First consider the
case where the manifold $\M$ is orientable.  We can subdivide the
entire boundary of the disk $\M\setminus G$ into six paths labeled
consecutively $\alpha, \beta, \gamma, \bar\beta, \bar\alpha, \delta$.
Here, $\bar\alpha$ and~$\bar\beta$ are the corresponding copies of
$\alpha$ and $\beta$ in the polygonal schema.  Because $\M$ is
orientable, $\alpha$ and $\bar\alpha$ have opposite orientations, as
do $\beta$ and $\bar\beta$.  Either or both of $\gamma$ and $\delta$
could be empty.  See the lower left part of Figure~\ref{F:cutpath}.
The subpath~$\sigma'$ can enter the interior of the disk $\M\setminus
G$ from four of these six paths ($\alpha$, $\beta$, $\bar\alpha$,
and~$\bar\beta$) and leave the interior of the disk through any of the
six paths.

\begin{figure}[t]
    \centerline{\includegraphics[width=\textwidth]{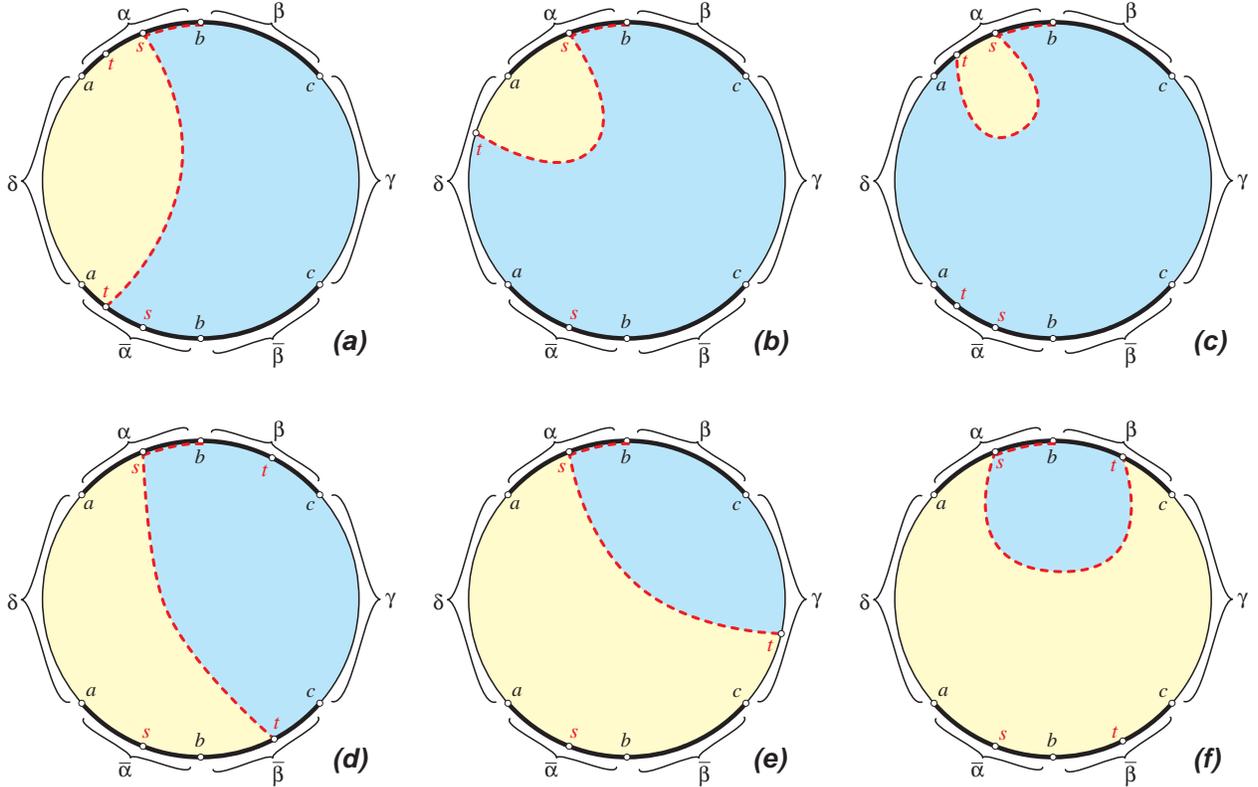}}
    \caption{Six cases for the proof of Lemma \ref{L:tight} for
       orientable manifolds; all other cases are reflections of
       these.  In each case, some subpath of~$\alpha$~or~$\beta$
       appears on the boundary of both sub-disks.}
    \label{F:cases}
\end{figure}

\def\concat{\cdot}

Suppose $\sigma'$ enters the interior of $\M\setminus G$ from
$\alpha$; the other three cases are symmetric.  Figure \ref{F:cases}
shows the six essentially different ways for $\sigma'$ to leave the
interior of $\M\setminus G$.  In each case, we easily verify that
after cutting along $\sigma'$, some subpath $\sigma$ of either
$\alpha$ or $\beta$ is on the boundary of both disks.  Specifically:

\begin{enumerate}[(i)]
\item
If $\sigma'$ leaves through $\alpha$ or $\bar\alpha$ (see
Figures~\ref{F:cases}(c) and~\ref{F:cases}(a), respectively), then
$\sigma$ is the subpath of $\alpha$ from $s$ to $t$.  Since both
$\sigma$ and $\sigma'$ have the same endpoints and $\sigma'$ is a
shortest path, $\sigma$ must be longer than $\sigma'$.

\item
If $\sigma'$ leaves through $\beta$ or $\bar\beta$ (see
Figures~\ref{F:cases}(f) and~\ref{F:cases}(d), respectively), then
$\sigma$ is the subpath of $\beta$ from $b$ to $t$.  Indeed, $\rho =
\alpha[b,s] \concat \sigma'$ is the shortest path from $b$ to $t$, and
as such $\abs{\sigma'} \leq \abs{\rho} < \abs{\beta[b,t]} =
\abs{\sigma}$.  (Here $\concat$ denotes path concatenation, and
$\alpha[x,y]$ denotes the subpath of $\alpha$ from $x$ to $y$.)

\item
If $\sigma'$ leaves through $\gamma$ (see Figure~\ref{F:cases}(e)), then
$\sigma = \beta$. Clearly, $\abs{\sigma'} < \abs{\alpha} = \abs{\beta} =
\abs{\sigma}$.
 
\item
Finally, if $\sigma'$ leaves through $\delta$ (see
Figure~\ref{F:cases}(b)), then $\sigma$ is the subpath of $\alpha$
from $a$ to~$s$.  Clearly, $\abs{\sigma'} < \abs{\alpha[s,a]} \leq
\abs{\sigma}$.

\end{enumerate}

If $\M$ is non-orientable, the path $\alpha\beta$ could appear either
with the same orientation or with opposite orientations on the
boundary of the disk $\M\setminus G$.  If the orientations are
opposite, the previous case analysis applies immediately.  Otherwise,
the boundary can be subdivided into six paths labeled consecutively
$\alpha, \beta, \gamma, \alpha, \beta, \delta$.  Without loss of
generality, $\sigma'$ enters the interior of $\M\setminus G$ from
$\alpha$ and leaves through any of these six paths.  The six cases are
illustrated in Figure \ref{F:cases2}.  Again, we easily verify that in
each case, some subpath $\sigma$ of either $\alpha$ or $\beta$ is on
the boundary of both disks.  We omit further details.
\end{proof}

For any cut graph $G$ of a manifold $\M$, we define the corresponding
\emph{reduced} cut graph~$\hat{G}$ as follows.  First we remove any
topologically trivial cuts; that is, we repeatedly remove any edge
with a vertex of degree $1$ that is not on the boundary of $\M$.  We
then augment the cut graph by adding all the boundary edges of $\M$.
Finally, we contract each maximal path through degree-$2$ vertices
into a single edge.  The resulting reduced cut graph $\hat{G}$ is
$2$-edge connected, and each of its vertices has degree at least $3$.
Every vertex of $\hat{G}$ is either a branch point or a boundary point
of $G$, and every edge of $\hat{G}$ corresponds to either a cut path
or a boundary path in $G$.  However, in general, not all branch points
and cut paths in $G$ are represented in $\hat{G}$.

\begin{lemma}
\label{L:small}
Let $\M$ be a polyhedral $2$-manifold with genus $g$ and $k$ boundary
components.  Any reduced cut graph $\hat{G}$ of $\M$ is connected, has
between $\max\set{1,k}$ and $4g+2k-2$ vertices, and has between
$g+\max\set{0,2k-1}$ and $6g+3k-3$ edges.
\end{lemma}

\begin{proof}
Let $G'$ be the cut graph corresponding to $\hat{G}$, after all the
trivial cuts have been removed.  The boundary of the polygonal schema
$\M\setminus G'$ can be partitioned into cut paths and boundary paths,
each corresponding to an edge in $\hat{G}$.  Thus, $\hat{G}$ is
connected.

Let $v$ and $e$ denote the number of vertices and edges in $\hat{G}$,
respectively.  If any vertex in~$\hat{G}$ has degree $d\ge 4$, we can
replace it with $d-3$ trivalent vertices and $d-3$ new edges of length
zero.  Thus, in the worst case, every vertex in $\hat{G}$ has degree
exactly $3$, which implies that $3v=2e$.  Since $\hat{G}$ is embedded
in $\M$ with a single face, Euler's formula implies that $v-e+1 = \chi
= 2-2g-k$ if $\M$ is orientable, and $v-e+1 = \chi = 2-g-k$ if $\M$ is
non-orientable.  It follows that $v \le 4g+2k-2$ and $e \le 6g+3k-3$,
as claimed.

On the other hand, $\hat{G}$ has at least one vertex on each boundary
component of $\M$, and at least one vertex even if $\M$ has no
boundary, so $v\ge \max\set{1, k}$.  Thus, Euler's formula implies
that $2-2g-k = v-e+1 \ge \max\set{1, k}-e+1$ if $\M$ is orientable, or
equivalently, $e \ge 2g + \max\set{0, 2k-1}$.  Similarly, if $\M$ is
non-orientable, Euler's formula implies that $e \ge g + \max\set{0,
2k-1}$.
\end{proof}

Our minimum cut graph algorithm exploits Lemma \ref{L:tight} by
composing potential minimum cut graphs out of $O(g+k)$ shortest paths.
Unfortunately, a single pair of nodes in $\M$ could be joined by
$2^{\Omega(n)}$ shortest paths, in $2^{\Omega(g+k)}$ different isotopy
classes, in the worst case.  To avoid this combinatorial explosion, we
can add a random infinitesimal weight $\e \cdot w(e)$ to each
edge~$e$.  The Isolation Lemma of Mulmuley, Vazirani, and
Vazirani~\cite{mvv-miemi-87} implies that if the weights $w(e)$ are
chosen independently and uniformly from the integer set $\set{1, 2,
\dots, n^2}$, all shortest paths are unique with probability at least
$1-1/n$; see also \cite{crs-rouei-95, ks-reitm-01}.%
\footnote{Alternately, if we choose $w(e)$ uniformly from the
   \emph{real} interval $[0,1]$, shortest paths are unique with
   probability~$1$.  This may sound unreasonable, but recall that no
   polynomial-time algorithm is known to compare sums of square roots
   of integers in any model of computation that does not include
   square root as a primitive operation~\cite{b-csrpt-91}.  Thus, to
   compute Euclidean shortest paths in a geometric graph with integer
   vertex coordinates, we must either assume exact real arithmetic or
   (grudgingly) accept some approximation error~\cite{ggj-ccsmt-77}.}

We are now finally ready to describe our minimum cut graph algorithm.

\begin{theorem}
The minimum cut graph of a polyhedral $2$-manifold $\M$ with genus
$g$ and $k$ boundary components can be computed in time
$n^{O(g+k)}$.
\end{theorem}

\begin{proof}
We begin by computing the shortest path between every pair of vertices
in $\M$ in $O(n^2\log n)$ time by running Dijkstra's single-source
shortest path algorithm for each vertex~\cite{d-ntpcg-59,j-easps-77},
breaking ties using random infinitesimal weights as described above.
Once these shortest paths and midpoints have been computed, our
algorithm enumerates by brute force every possible cut graph that
satisfies Lemmas~\ref{L:tight} and~\ref{L:small}, and returns the
smallest such graph.

Each cut graph is specified by a set $V$ of up to $4g+2k-2$ vertices
of $\M$, a set $E$ of up to $6g+3k-3$ edges of $\M$, a trivalent
multigraph $\hat{G}$ with vertices $V$, and a assignment of edges in
$E$ to edges in $\hat{G}$.  Each edge $(v,w)$ of $\hat{G}$ is assigned
a unique edge $e\in E$ to define the corresponding cut path in $\M$.
This cut path is the concatenation of the shortest path from $v$ to
$e$, $e$ itself, and the shortest path from $e$ to $w$.  If the
midpoint of this cut path is not in the interior of~$e$, we declare
the cut path invalid, since it violates Lemma \ref{L:tight}.  (Because
shortest paths between vertices are unique, the midpoint of any cut
path in the minimal cut graph must lie in the interior of an edge.)
If all the cut paths are valid, we then check that every pair of cut
paths is disjoint, except possibly at their endpoints, and that
removing all the cut paths from $\M$ leaves a topological disk.

Our brute-force algorithm considers $O(n^{4g+2k-2})$ different vertex
sets~$V$, $O(n^{6g+3k-2})$ different edge sets~$E$, at most
$\binom{(4g+2k-2)^2}{6g+3k-2}$ different graphs~$\hat{G}$ for each
vertex set, and at most $(6g+3k-2)!$ different assignments of edges in
each graph $\hat{G}$ to edges in each edge set $E$.  Thus,
$n^{O(g+k)}$ potential cut graphs are considered altogether.  The
validity of each potential cut graph can be checked in $O(n)$ time.
\end{proof}

% ----------------------------------------------------------------------
\section{Finding Short Nontrivial Cycles}

As a step towards efficiently computing approximate minimum cut
graphs, we develop algorithms to compute shortest and nearly-shortest
nontrivial cycles in arbitrary $2$-manifolds, possibly with boundary.
Although our most efficient approximation algorithm for cut graphs
requires only approximate shortest nontrivial cycles on manifolds
without boundary, we believe these algorithms are of independent
interest.

\begin{figure}
\centerline{\includegraphics[height=1.5in]{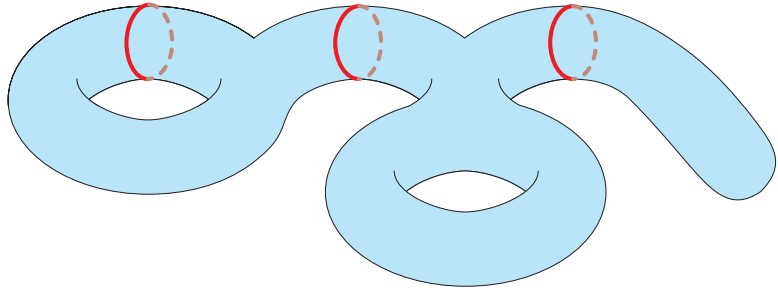}}
\caption{From left to right: non-separating, essential but separating,
and trivial cycles on a $2$-manifold.}
\label{Fig:cycles}
\end{figure}

We distinguish between two type of nontrivial simple cycles.  A simple
cycle $\gamma$ in $\M$ is \emph{non-separating} if $\M\setminus\gamma$
has only one connected component.  A simple cycle $\gamma$ in $\M$ is
\emph{essential} if it is not contractible to a point or a single
boundary cycle of $\M$.  Every non-separating cycle is essential, but
the converse is not true.  Formally, non-separating cycles are
\emph{homologically} nontrivial, and essential cycles are
\emph{homotopically} nontrivial.  See Figure \ref{Fig:cycles}.

As can be seen immediately from Figure \ref{Fig:cycles}, it is not
possible to determine whether a cycle is non-separating, essential, or
trivial by examining only a local neighborhood.  Dey and Schipper
\cite{ds-ntcps-95} describe an algorithm to determine whether an
arbitrary circuit is contractible in $O(n)$ time; their algorithm
begins by computing an arbitrary cut graph of the manifold.
Fortunately, as we shall see shortly, we can simplify their algorithm
considerably when the given cycle is simple.

\subsection{Shortest Cycles}

We begin by describing how to find the shortest nontrivial cycle
through a given vertex.  Our algorithm uses a combination of
Dijkstra's single-source shortest path algorithm~\cite{d-ntpcg-59} and
a modification of the canonical polygonal schema algorithm of Lazarus
\etal~\cite{lpvv-ccpso-01}.  Our algorithm is similar to the approach
taken by Guskov and Wood \cite{gw-tnr-01} to find and remove small
handles from geometric models reconstructed from noisy data; see
also~\cite{whds-its-02}.

The algorithm of Lazarus \etal\ builds a connected subset~$S$ of a
triangulated manifold without boundary, starting with a single
triangle and adding new triangles on the boundary of $S$ one at a
time.  If a new triangle intersects the boundary of $S$ in more than
one component, the algorithm checks which of the following three cases
holds: (1)~$\M\setminus S$ is connected; (2)~neither component of
$\M\setminus S$ is a disk; or (3)~one component of $\M\setminus S$ is
a disk.  In the final case, the algorithm adds the disk component to
$S$ and continues searching the other component of $\M\setminus S$.
If we run this algorithm until either case (1) or case~(2) holds, the
total running time is $O(n)$.  See Lazarus \etal~\cite{lpvv-ccpso-01}
for further details.

First we describe a straightforward generalization of the algorithm
used by Lazarus to determine the structure of $\M\setminus S$.

\begin{lemma}
\label{L:classify}
Let $\M$ be a connected polyhedral $2$-manifold~$\M$ whose genus $g$
and number of boundary components $k$ are known.  Any simple cycle
$\gamma$ in $\M_1$ can be classified as non-separating, essential but
separating, or trivial in $O(m)$ time, where $m$ is the complexity of
the smaller component of $\M\setminus\gamma$.  In particular, if
$\gamma$ is non-separating, the running time is $O(n)$.
\end{lemma}

\begin{proof}
We perform two simultaneous depth-first searches, starting from either
side of any vertex of $\gamma$.  If either search meets a vertex
already visited by the other, $\gamma$ is a non-separating cycle.  The
running time in this case is trivially $O(n)$.

Conversely, if one search halts without meeting the other, $\gamma$ is
a separating cycle.  Let~$C$ be the smaller component of $\M \setminus
\gamma$.  To determine whether $\gamma$ is essential, we compute the
Euler characteristic and the number of boundary components of $C$,
which we denote by $\chi(C)$ and $k(C)$, respectively.  (This can be
done on the fly during the depth-first search phase.)  Let $\chi$
denote the Euler characteristic of $\M$.  The cycle $\gamma$ is
contractible if and only if one of the following conditions holds.
\begin{itemize}
\item
$C$ is a disk, or equivalently, $\chi(C) = 1$.

\item
$C$ is an annulus, or equivalently, $\chi(C) = 0$ and $k(C) = 2$.

\item
$\M\setminus C$ is a disk, or equivalently, $\chi(C) = \chi - 1$.  The
equivalence follows from the inclusion-exclusion formula $\chi(A\cup
B) = \chi(A) + \chi(B) - \chi(A\cap B)$.

\item
$\M\setminus C$ is an annulus, or equivalently, $\chi(C) = \chi(\M)$
and $k(C) = k(M)$.

\end{itemize}
If none of these conditions hold, then $\gamma$ is essential.  The
total running time if $\gamma$ is a separating cycle is $O(m)$, where
$m$ is the complexity of $C$.
\end{proof}

Note that we can simplify this algorithm slightly if the manifold has
no boundary, since in that case neither component of $\M \setminus
\gamma$ is an annulus.  We will use this simplification in our
approximate cut graph algorithm.

\begin{lemma}
\label{L:one-essential}
Let $u$ be a vertex of a polyhedral $2$-manifold~$\M$, possibly with
boundary.  The shortest essential cycle in $\M_1$ that contains $u$
can be computed in $O(n\log n)$ time.
\end{lemma}

\begin{proof}
We find the shortest essential cycle through~$u$ by simulating a
circular wave expanding from $u$.  Whenever the wave touches itself,
either we have the shortest essential cycle through $u$, or one
component of the wave bounds a disk in $\M$ and we can continue
expanding the other component.

We modify the algorithm of Lazarus \etal~\cite{lpvv-ccpso-01} as
follows.  First, $S$ is no longer a set of triangles but a more
general connected subset of vertices, edges, and facets of $\M$.
Initially, $S$ contains only the source vertex~$u$.  Second, we use
Dijkstra's algorithm to determine the order for edges to be added.  We
add a facet to $S$ only when all its vertices have been added to $S$,
either directly or as part of another facet.  We run the algorithm
from Lemma \ref{L:classify} whenever $S$ is no longer simply
connected, that is, when we add a new edge $vw$ with both endpoints on
the boundary of~$S$.  If $\M\setminus S$ is disconnected, we continue
only if one component of $\M\setminus S$ is a disk \emph{or an
annulus}, as can be checked using Lemma \ref{L:classify}.  In that
case, we add the disk or annulus component of $\M\setminus S$ to $S$,
discard the vertices of that component from the Dijkstra priority
queue, and continue searching in the other component.  If $\M\setminus
S$ is connected, or if neither component is a disk or an annulus, we
have found the shortest essential cycle through~$u$, consisting of the
shortest path from~$u$ to~$v$, the edge $vw$, and the shortest path
from $w$ to~$u$.

Dijkstra's algorithm requires $O(n\log n)$ time.  Each time we find a
trivial cycle, we spend $O(m)$ time and discard a disk with complexity
at least $m$.  Thus, the total time spent performing cycle
classification and maintaining the wavefront set $S$ is $O(n)$.  Thus,
the total running time of our algorithm is $O(n\log n)$.
\end{proof}

Running this algorithm once for every vertex of $\M$ immediately gives
us the shortest essential cycle.

\begin{corollary}
Let $\M$ be a polyhedral 2-manifold, possibly with boundary.  The
shortest essential cycle in $\M_1$ can be computed in $O(n^2\log n)$
time.
\end{corollary}

A simple modification of our algorithm allows us to find shortest
\emph{non-separating} cycles in the same asymptotic time.

\begin{lemma}
\label{L:one-nonsep}
Let $u$ be a vertex of a polyhedral $2$-manifold~$\M$, possibly with
boundary.  The shortest non-separating cycle in $\M_1$ that contains
$u$ can be computed in $O(n\log n)$ time.
\end{lemma}

\begin{proof}
The only change from the previous algorithm is that if we discover an
essential separating cycle, we continue recursively in both components
of $\M\setminus S$.  The cost of Dijkstra's algorithm is still
$O(n\log n)$, but we now must spend extra time in the
cycle-classification algorithm of Lemma~\ref{L:classify}.  As before,
the total time spent finding trivial cycles is $O(n)$, since we can
charge the search time to the discarded components.

Let $T(n, g)$ denote the total time spent finding separating essential
cycles.  This function satisfies the recurrence
\[
	T(n, g) \le T(m, h) + T(n-m, g-h) + O(m),
\]
where $m\le n/2$ is the complexity of the smaller component of
$\M\setminus S$ and $h$ is its genus.  The base case of the recurrence
is $T(n, 1) = 0$, since every essential cycle on a genus-$1$ surface
is non-separating.

Similar recurrences appear in the analysis of output-sensitive planar
convex hull algorithms \cite{bs-spoos-97, csy-pddpo-97, ks-upcha-86,
w-rqh-97}, suggesting that the solution to our recurrence is $T(n,g) =
O(n\log g)$.  Indeed, we can prove this by induction as follows.
Suppose
\[
	T(n,g)  \le T(m, h) + T(n-m, g-h) + cm
\]
for some constant $c$.  We claim that $T(n,g) \le cn\lg g$.  The
inductive hypothesis implies that
\begin{align*}
	T(n,g)  &\le cm\lg h + c(n-m)\lg(g-h) + cm
\\		&\le \max_{1\le h\le g-1}
			\left(cm \lg h + c(n-m)\lg (g-h)\right) + cm.
\end{align*}
A simple application of derivatives implies that the right hand side
of this inequality is maximized when $h = mg/n$.  Thus,
\begin{align*}
	T(n,g)	&\le cm\lg\frac{mg}{n} +
				c(n-m)\lg\frac{(n-m)g}{n} + cm
\\		&= cn\lg g + cm\lg m + c(n-m)\lg (n-m) - cn\lg n + cm.
\end{align*}
Since $m\le n/2$ and $n-m\le n$, we can simplify this inequality to
\[
	T(n,g)	\le cn\ln g + cm\lg (n/2) + c(n-m)\lg n - cn\lg n + cm
		= cn\lg g,
\]
completing the proof.

Thus, the total time spent in the cycle-classification phase of our
algorithm is $O(n\log g)$.  Since $g\le n$, this is dominated by the
cost of maintaining the Dijkstra priority queue.
\end{proof}

\begin{corollary}
Let $\M$ be an polyhedral 2-manifold, possibly with boundary.  The
shortest non-separating cycle in $\M_1$ can be computed in $O(n^2\log
n)$ time.
\end{corollary}

\subsection{Nearly-Shortest Cycles}

As we will argue in the next section, computing short nontrivial
cycles is the bottleneck in our approximate cut graph algorithm.
Fortunately, exact minimum cycles are not necessary for our results.
We can speed up our cut graph algorithm, without significantly
increasing the approximation factor, by searching for a nontrivial
cycle at most twice as long as the shortest.  Our approximation
algorithm assumes that the manifold $\M$ has no boundary; fortunately,
as we shall see in the next section, this is sufficient for our
purposes.

Our approximation algorithm works as follows.  First, we compute a set
of shortest paths (in fact, a cut graph) that intersects every
non-separating cycle in the manifold $\M$.  Then we contract each
shortest path $\pi$ in this set to a point and find the shortest
nontrivial cycle through that point, as described by
Lemmas~\ref{L:one-essential} and~\ref{L:one-nonsep}.

\begin{lemma}
\label{L:shortest-path}
Let $\pi$ be a shortest path between two vertices in a polyhedral
$2$-manifold~$\M$, and let $\gamma^*$ be the shortest essential
(resp.\ non-separating) cycle in $\M_1$ that intersects $\pi$.  In
$O(n\log n)$ time, one can compute an essential (resp.\
non-separating) cycle $\gamma$ in $\M$ such that $\abs{\gamma} \le 2
\abs{\gamma^*}$.
\end{lemma}

\begin{proof}
Let $\M'$ be the manifold obtained by contracting the shortest path
$\pi$ to a single vertex $v$.  Because $\pi$ has no cycles, $\M'$ has
the same topological type as $\M$.  Let $\gamma'$ be the shortest
essential (resp.\ non-separating) cycle in $\M'$ that passes through
$v$.  Clearly, $\abs{\gamma'} \le \abs{\gamma^*}$.  We can compute
this cycle in $O(n\log n)$ time by Lemma \ref{L:one-essential}
(resp.\ Lemma \ref{L:one-nonsep}).

We construct a cycle $\gamma$ in $\M$ by concatenating two paths
$\alpha$ and $\beta$, where $\alpha$ contains the edges of $\gamma'$
and $\beta$ is the subpath of $\pi$ between the endpoints of $\alpha$.
The sequence of edge contractions that transforms $\M$ to $\M'$ also
transforms $\gamma'$ to $\gamma$.  Hence, $\gamma'$ is an essential
cycle of $\M$.  Because $\beta$ is a subpath of a shortest path,
$\beta$ is actually the shortest path between the endpoints of
$\alpha$, so $\abs{\beta} \le \abs{\alpha} = \abs{\gamma'}$.  It
follows that $\abs{\gamma} = \abs{\alpha} + \abs{\beta} \le
2\abs{\gamma'} \le 2\abs{\gamma^*}$.
\end{proof}

This lemma suggests a natural algorithm for finding a short nontrivial
cycle: Compute a set of shortest paths that intersect every
non-separating cycle (and thus every essential cycle), and then run
the algorithm from Lemma \ref{L:shortest-path} for each path in this
set.

\begin{lemma}
\label{L:shortcut}
Let $\M$ be a polyhedral $2$-manifold without boundary.  In $O(n\log
n)$ time, one can compute a set $\Pi$ of $O(g)$ shortest paths on
$\M_1$ such that every non-separating cycle (and thus every essential
cycle) in $\M_1$ intersects at least one path in $\Pi$.
\end{lemma}

\begin{proof}
We compute a cut graph $G$ as follows.  First we compute a
shortest-path tree $T$ from an arbitrary initial vertex $v$ using
Dijkstra's algorithm.  We then compute an arbitrary spanning tree
$T^*$ of the dual of $\M\setminus T$, that is, the graph whose
vertices are facets of $\M$ and whose edges join pairs of facets that
share a common edge \emph{not} in $T$.  Analysis similar to
Lemma~\ref{L:small} implies that there are $O(g)$ edges that do not
appear in $T$ and whose dual edges to not appear in $T^*$.  Call this
set of unclaimed edges $E$.  Let $\Pi$ be the set of $O(g)$ shortest
paths from $v$ to the endpoints of~$E$; these paths are all in $T$.
Finally, let $G = \Pi \cup E'$.

We easily observe that $\M\setminus G$ is a topological disk, so $G$
is a cut graph.  It follows that every non-separating cycle in $\M_1$
intersects $G$.  Since every vertex of $G$ is also a vertex of some
path in~$\Pi$, every non-separating cycle in $\M_1$ intersects at
least one path in $\Pi$.
\end{proof} 

Notice that this algorithm does not work if $\M$ has a boundary, since
the dual graph of $\M\setminus T$ could be disconnected.

\begin{corollary}
\label{C:fast-nontriv}
Let $\M$ be a polyhedral $2$-manifold with genus $g$ and no boundary,
and let~$\gamma^*$ be its shortest essential (resp.\ non-separating)
cycle.  In $O(gn\log n)$ time, one can compute an essential (resp.\
non-separating) cycle~$\gamma$ in~$\M_1$ such that $\abs{\gamma} \le
2\abs{\gamma^*}$.
\end{corollary}

\begin{proof}
We construct a set $\Pi$ of $O(g)$ shortest paths, at least one of
which is guaranteed to intersect $\gamma^*$, as described in the
previous lemma.  Then for each path $\pi\in\Pi$, we contract $\pi$ to
a point and find the shortest nontrivial cycle through that point in
$O(n\log n)$ time, as described by Lemma~\ref{L:shortest-path}.
\end{proof}

% ----------------------------------------------------------------------
\section{Approximate Minimum Cut Graphs}

We now describe a simple polynomial-time greedy algorithm to construct
an approximate minimum cut graph for any polyhedral manifold~$\M$.

To handle manifolds with boundary, it will be convenient to consider
the following simplified form.  Given a manifold $\M$ with genus $g$
and $k$ boundary components, the corresponding \emph{punctured}
manifold $(\MM, P)$ consists of a manifold~$\MM$ with the same genus
as $\M$ but without boundary, and a set $P$ of $k$ points in~$\MM$,
called \emph{punctures}.  To construct $\MM$, we contract every
boundary component of $\M$ to a single point, which becomes one of the
punctures in~$P$.%
\footnote{We could simulate this contraction by artificially assigning
every boundary edge of~$\M$ a weight of zero, although this would
require a few simple changes in our algorithms.}
If any vertex of $\M$ has multiple edges to the same boundary
component, $\MM$ contains only the edge with smallest weight, breaking
ties using the Isolation Lemma as above.  If $\M$ has no boundary,
then $\MM = \M$ and $P = \varnothing$.

Our goal now is to compute the minimum cut graph of $\MM$ that touches
every puncture in $P$; henceforth, we call this simply the minimum cut
graph of $(\MM, P)$.  This reduction is motivated by the following
trivial observation.

\begin{lemma}
The minimum cut graph of any polyhedral $2$-manifold $\M$ has the same
length as the minimum cut graph of $(\MM, P)$.
\end{lemma}

Our approximation algorithm works as follows.  We repeatedly cut along
short nontrivial cycles until our surface becomes a collection of
punctured spheres, connect the punctures on each component by cutting
along a minimum spanning tree, and finally (if necessary) reglue some
previously cut edges to obtain a single disk.  The resulting cut graph
is composed of a subset of the edges of the short nontrivial cycles
and all the edges of the minimum spanning forest.

\subsection{Using Short Nontrivial Cycles}

The first component of our algorithm is a subroutine to compute
approximately shortest nontrivial cycles, described by
Corollary~\ref{C:fast-nontriv}.  As required by that algorithm, the
input manifold $\MM$ has no boundary; the punctures are completely
ignored.

The following argument relates the length of the shortest nontrivial
cycles to the length of the minimum cut graph.

\begin{lemma}
\label{L:cyclefactor}
Let $G$ be any cut graph of a polyhedral $2$-manifold $\MM$ with genus
$g$ and no boundary.  The shortest cycle in $G$ contains $O((\log
g)/g)$ of the total length of $G$.
\end{lemma}

\begin{proof}
First consider the reduced cut graph $\hat{G}$, constructed by
repeatedly contracting any edge with a vertex of degree less than
three, as in Section~\ref{S:algorithm}.  Every vertex in $\hat{G}$ has
degree at least $3$.  Without loss of generality, assume that every
vertex in $\hat{G}$ has degree exactly $3$, splitting each high-degree
vertex into a tree of degree-$3$ vertices if necessary, as in the
proof of Lemma~\ref{L:small}.  A straightforward counting argument
implies that any trivalent graph whose girth (minimum cycle length) is
$c$ must have at least $3\sqrt{2} \cdot 2^{c/2} - 2$ vertices if $c$
is odd, and at least $2\cdot 2^{c/2} -2$ vertices if $c$ is
even~\cite{b-ccglg-98}.  By Lemma~\ref{L:small}, $\hat{G}$ has at most
$4g-2$ vertices, so $\hat{G}$ must have a cycle $\hat{\gamma}$ with
most $2(\lg g+1) = O(\log g)$ edges.

Starting with $\hat{G}_0 = \hat{G}$, we inductively define a sequence
of reduced graphs $\hat{G}_1, \hat{G}_2, \dots$ as follows.  For each
$i>0$, let $\hat{\gamma}_i$ denote the shortest cycle in
$\hat{G}_{i-1}$.  We obtain $\hat{G}_i$ by reducing the graph
$\hat{G}_{i-1}\setminus \hat{\gamma}_i$, or equivalently, removing the
vertices of $\hat{\gamma}_i$ and all their edges, and then contracting
$\abs{\hat{\gamma}_i}$ nearby length-$2$ paths to single edges.  Our
earlier argument implies that each cycle $\hat{\gamma}_i$ has at most
$2(\lg g + 1)$ edges.  Thus, for each $i$, we have $\abs{E(\hat{G}_i)}
= \abs{E(\hat{G}_{i-1})} - 6(\lg g + 1)$.  Lemma \ref{L:small} implies
that the original reduced cut graph $\hat{G}$ has at least $g$ edges,
so we can repeat this process at least $g/6(\lg g + 1)$ times.

Let $\gamma_i$ denote the cycle in the original cut graph $G$
corresponding to $\hat{G}_i$.  By our construction, $\gamma_i$ and
$\gamma_j$ are disjoint for all $i\ne j$, so we have a set of at least
$g/6(\lg g + 1)$ disjoint cycles in $G$.  At least one of these cycles
has length at most $6(\lg g + 1)/g = O((\log g)/g)$ times the total
length of~$G$.
\end{proof}

Since every cycle in the minimum cut graph is non-separating, and
therefore essential, we immediately have the following corollary.

\begin{corollary}
For any polyhedral $2$-manifold $\MM$ with genus $g$ and no boundary,
both the length of the shortest non-separating cycle and the length of
the shortest essential cycle are at most $O((\log g)/g)$ times the
length of the minimum cut graph of $\MM$.
\end{corollary}

\subsection{Puncture-Spanning Trees}

The second component of our cut graph algorithm is a subroutine to
compute the \emph{minimum puncture-spanning tree} of a punctured
manifold $(\MM, P)$, that is, the minimum spanning tree of the
punctures~$P$ in the shortest-path metric of $\MM_1$.

\begin{lemma}
\label{L:mstalgo}
The minimum puncture-spanning tree of any punctured polyhedral
$2$-manifold $(\MM, P)$ can be computed in $O(n \log n)$ time.
\end{lemma}

\begin{proof}
We simulate Kruskal's minimum spanning tree algorithm
\cite{k-sssgt-56, t-dsna-83} by adding shortest puncture-to-puncture
paths one at a time, in increasing order of length.  To compute the
shortest paths, we simultaneously propagate wavefronts from all $k$
punctures using Dijkstra's algorithm.  Whenever two wavefronts (\ie,
two growing shortest-path trees) collide, we add a new edge to the
evolving minimum spanning tree and merge those two wavefronts.  To
implement this algorithm efficiently, we maintain the wavefronts in a
union-find data structure.  The resulting running time is $O(n\log
n)$.
\end{proof}

This is essentially the algorithm proposed by Takahashi and
Matsuyama~\cite{tm-asnst-80} to compute approximate Steiner trees in
arbitrary graphs.  The same algorithm was also recently used by
Sheffer \cite{s-stsrp-02} and by L\'evy \etal~\cite{lprm-lscma-02} to
compute cut graphs, where surface features with high discrete
curvature play the role of punctures.

\begin{lemma}
\label{L:mstfactor}
The length of the minimum puncture-spanning tree of any punctured
polyhedral $2$-manifold $(\MM, P)$ is at most twice the length of any
cut graph of $(\MM, P)$.
\end{lemma}

\begin{proof}
The \emph{minimum Steiner tree} of $P$ is the subgraph of $\MM_1$
of minimum total weight that includes every point in~$P$.  Since
any cut graph of $(\MM, P)$ must touch every puncture, no cut
graph is shorter than this minimum Steiner tree.  On the other
hand, the minimum spanning tree of $P$ has at most twice the
length of the minimum Steiner tree~\cite{kmb-fast-81,tm-asnst-80}.
\end{proof}

\subsection{Analysis}

We now have all the components of our greedy cut graph algorithm.  At
any stage of the algorithm, we have a punctured manifold $(\MM, P)$.
Our algorithm repeatedly cuts along a short non-separating cycle of
$\MM$, using Corollary \ref{C:fast-nontriv}.  This cut creates one or
two new boundary circles, which we collapse to new punctures.  When
the manifold is reduced to a collection of punctured spheres, we cut
along the minimum puncture-spanning tree of each component using the
algorithm in Lemma~\ref{L:mstalgo}.

Each non-separating cycle cut reduces the genus of $\M$ by~$1$.  This
immediately implies that our algorithm performs exactly $g$ cycle
cuts, so the overall running time is
\[
	g \cdot O(g n\log n) + O(n \log n) = O(g^2 n\log n).
\]

For any graph $X$, let $\abs{X}$ denote its total length.  Let $G^*$
denote the minimum cut graph of $(\MM, P)$.  Let $(\MM_i, P_i)$ denote
the punctured manifold after $g-i$ cycle cuts have been performed, so
$\M_i$ has genus~$i$, and let $G^*_i$ denote the minimum cut graph of
$\MM_i$, \emph{ignoring} the punctures $P_i$.  Since collapsing edges
cannot increase the minimum cut graph length, we have $\abs{G^*_i} \le
\abs{G^*_g} \le \abs{G^*}$ for all $i$.

Let $\gamma_i$ denote the short non-separating cycle of $\M_i$ found
by Corollary~\ref{C:fast-nontriv}.  (We easily observe that \emph{any}
cut graph of $\MM_i$ must intersect this cycle.)
Lemma~\ref{L:cyclefactor} and Corollary~\ref{C:fast-nontriv} imply
that $\abs{\gamma_i} \le O((\log i)/i)\cdot \abs{G^*_i}$ for all $i$
and $j$.  Summing over all $g$ cuts, we conclude that the total length
of all cycle cuts is at most
\[
	\sum_{i=1}^g O((\log i)/i) \cdot \abs{G_i^*}
	=
	O(\log^2 g) \cdot \abs{G^*}.
\]

Similarly, Lemma \ref{L:mstfactor} implies that the minimum
puncture-spanning forest has length at most $2\abs{G^*}$.  Finally,
regluing previously cut edges to obtain a single disk only reduces the
length of the final cut graph.  Thus, the final cut graph computed by
our algorithm has length at most $O(\log^2 g)\cdot \abs{G^*}$.

\begin{theorem}
\label{Th:approx}
Given a polyhedral $2$-manifold $\M$ with genus $g$ and $k$ boundary
components, an $O(\log^2 g)$-approximation of its minimum cut graph
can be constructed in $O(g^2n\log n)$ time.
\end{theorem}

Cutting along short essential cycles instead of short non-separating
cycles leads to exactly the same asymptotic running time and
approximation bounds, although the algorithm and its analysis are
slightly more complicated.  For purposes of analysis, we can divide
the algorithm into phases, where in the $i$th phase, we cut along a
short essential cycle of every component of the manifold that has
genus $i$.  Essential cycle cuts can separate the manifold into
multiple components, but since each component must have nontrivial
topology, the algorithm performs at most $g-1$ separating cuts.  At
the end of the algorithm, if necessary, we reglue along some
previously cut edges to obtain a single topological disk.  We refer to
the earlier version of this paper for further
details~\cite{eh-ocsd-02}.

\section{Open Problems}

We have developed new algorithms to compute exact and approximate
minimal cut graphs for manifold surfaces with arbitrary genus and
arbitrary boundary complexity.  Our approximation algorithm is
particularly simple.

Our results suggest several open problems, the most obvious of which
is to improve the running times and approximation factors of our
algorithms.  Is the minimum cut graph problem \emph{fixed-parameter
tractable}~\cite{df-pc-99}?  That is, can we compute exact minimum cut
graphs in time $f(g,k)\cdot n^{O(1)}$ for some function $f$?  The
similarity to the Steiner problem offers some hope here, since the
minimum Steiner tree of $k$ nodes in an $n$-node graph can be computed
in $O(3^k n + 2^k n^2 + n^3)$ time~\cite{dw-spg-71, hw-cpcr-94}.

The approximation algorithm of Theorem~\ref{Th:approx} is somewhat
indirect.  It computes a short cut graph by repeatedly computing a
`reasonable' cut graph and then extracting a short nontrivial cycle
that interacts with this cut graph.  It is natural to conjecture that
one can compute such a short cut graph directly, resulting in a faster
algorithm.  In particular, we conjecture that an approximately minimum
cut graph can be computed in $O(g n\log n)$ time.

How well can we approximate the minimum cut graph in nearly-linear
time?  There are several simple heuristics to compute 'good' cut graphs
in $O(n\log n)$ time, such as the dual shortest-path algorithm used by
Gu \etal~\cite{ggh-gi-02} and by Steiner and Fischer
\cite{sf-c3dfo-02}, and the algorithm described in the proof of
Lemma~\ref{L:shortcut}.  How well do these algorithms approximate the
minimum cut graph?

More generally, is there a simple, practical, $O(1)$-approximation
algorithm, like the minimum spanning tree approximation of Steiner
trees?  In fact, it might be that our algorithm provides such an
approximation, as our current analysis seems to be far from tight.
Unfortunately, the general Steiner tree problem is
MAXSNP-hard~\cite{bp-spel1-89}, so an efficient $(1+\e)$-approximation
algorithm for arbitrary $\e>0$ seems unlikely.

Several authors have pointed out apparent tradeoffs between the
quality of parameterizations and the length of the required cut graph;
see, for example, Sorkine \etal~\cite{scgl-bdpmp-02}.  How hard is it
to compute the (approximately) shortest cut graph required for a
parameterization whose distortion is less than some given limit?
Conversely, how hard is it to (approximately) minimize the distortion
of a parameterization, given an upper bound on the permitted length of
the surface cuts?  The complexity of these problems clearly depends
the which distortion measure is used, but we expect almost any variant
of this problem to be NP-hard.

Finally, can our ideas be applied to other useful families of curves
on manifolds, such as homology generators (families of $2g$ cycles
that intersect in $g$ pairs) and pants decompositions (maximal sets of
pairwise disjoint essential cycles~\cite{t-tdgt1-97})?

\medskip
\noindent
\textbf{Acknowledgments.}  We would like to thank Herbert Edelsbrunner
for an enlightening initial conversation.  We are also grateful to
Noga Alon, Steven Gortler, John Hart, Benjamin Sudakov, Kim
Whittlesey, and Zo\"e Wood for helpful discussions.

% ----------------------------------------------------------------------
%  Sariel -- To change bibliography style to alpha, do this:
%		cp schema.bbl-sariel schema.bbl
%  The alpha style doesn't understand the url field, which is crucial
%  for some of the references.  I generated your bbl file using the
%  alpha style and added the URLs by hand. -- Jeff
% ----------------------------------------------------------------------
\bibliographystyle{newabuser}
\bibliography{schema,geom}

\end{document}